# What drives performance in molecular MPNNs? An operator-level factorial benchmark

Panyu Jiao[1,#], Shuizhou Chen[2,#], Yiheng Shen[1,#], Yuyang Wang[1], Runhai Ouyang[3], Wei Xie[1,*]

[1]Materials Genome Institute, Shanghai University, Shanghai 200444, China

[2]School of Computer Engineering and Science, Shanghai University, Shanghai 200444, China

[3]School of Materials Science and Engineering, Tongji University, Shanghai 201804, China

## Abstract

Message-passing neural networks (MPNNs) are widely used for molecular property prediction, but their deployment as monolithic architectures makes it difficult to identify how specific message-passing operators affect performance. We present an operator-level factorial benchmark that decomposes 2D molecular MPNNs into the three families of message-seed initialization, node-edge fusion, and node update operators. The resulting 84 configurations are benchmarked on ten MoleculeNet datasets under a shared experimental setup and statistical analysis protocol. Across this controlled design, performance variation is associated primarily with message construction rather than update complexity. Message-seed initialization shows significant family-level effects for both regression and classification, node-edge fusion shows a significant family-level effect for regression with descriptive advantages for concatenation-based mixing, and the update family shows no statistically supported effect for either endpoint family. A representation probe into the Quinethazone molecule further demonstrates that concatenation-based mixing can better differentiate chemically distinct heteroatoms and withstand oversmoothing than Hadamard gating. Representative configurations selected separately for classification and regression recover competitive performance relative to established molecular graph neural network (GNN) baselines, ranking numerically best on eight of ten benchmark datasets. These empirical results are interpreted through concise mechanistic analyses of representative node-edge fusion and update operators. Our findings provide empirical design heuristics for molecular MPNNs by turning model design from a search over monolithic architectures into a targeted assessment of where and how chemical information enters the message-passing pipeline.

# 1. Introduction

Molecular representation learning now spans attention-based graph transformers, hybrid or multimodal architectures, and geometry-aware three-dimensional equivariant models built on graph neural networks (GNNs)[1–5]. Despite such architectural expansion, two-dimensional (2D) topological message-passing neural networks (MPNNs) remain the practical workhorse and routine baseline of molecular property prediction because they are chemically intuitive, flexible, and effective across benchmark suites[6–10]. In practice, named 2D molecular architectures such as DMPNN[10] and AttentiveFP[11] are often applied across heterogeneous molecular tasks, encouraging a tacit expectation that one broadly good MPNN recipe can serve many different endpoint types.

That expectation, however, is difficult to justify because molecular endpoints are highly diverse. Solubility, bioactivity, toxicity, and quantum-chemical targets depend on different chemical signals, operate on different scales, and may reward different inductive biases[9]. A key practical partition also exists between regression and classification tasks: continuous property prediction and binary or multilabel activity prediction need not favor the same message-passing operators. If these endpoint-family differences enter through specific stages of message passing rather than through end-to-end model capacity alone, then operator-level rather than whole-model comparison is the right diagnostic. This perspective also refines how molecular MPNNs should be situated relative to the broader GNN methodology literature. Aggregation and aggregation-update mappings have received sustained formal attention, because they control expressive power, neighborhood scaling, and oversmoothing[12–15]. In practical molecular MPNNs, however, aggregation has often converged on summation as the standard permutation-invariant neighborhood reducer. This apparent consensus on the aggregation operator does not eliminate a more chemically specific design question: what information should be encoded in each pairwise message before aggregation occurs? Molecular graphs differ from many general graph benchmarks in that their edges are chemically informative objects rather than merely adjacency indicators. Bond order, aromaticity, conjugation, ring membership, and stereochemistry can all modulate how the state of one atom should influence another[7,10]. Consequently, the construction of atom–bond messages before summation remains an under-isolated source of architectural variation in molecular MPNNs[16].

The difficulty is that message construction and update are usually bundled with broader architectural changes. Directed bond-centered messaging, dynamic edge-conditioned filters, graph-network state variables, attention mechanisms, geometric terms, and readout choices often change together in models and formalisms such as DMPNN, ECC, MEGNet, SchNet, crystal graph networks, EGNN, and Graphormer[1–3,10,17–19]. When one full architecture outperforms another, it is therefore unclear whether the gain comes from the operator itself, from interactions among operators, or from the surrounding design and training protocol. Whole-model comparisons alone leave unresolved which operator families are associated with molecular MPNN performance, whether regression and classification endpoints favor different message-construction choices, whether expressive node updating matters empirically after messages have already been constructed and aggregated, and how much of the apparent advantage of specialized architectures can be recovered within a simpler MPNN design space.

In this study we address these questions through operator-level factorial benchmarking, in the spirit of broader GNN design-space evaluation[20]. We decompose 2D molecular message passing into message-seed initialization, node-edge fusion, and node update; hold sum aggregation, sum readout, and Chemprop-style featurization fixed; and evaluate 84 operator combinations across ten MoleculeNet tasks under shared scaffold splits and matched tuning. The full space is screened with paired standardized statistics, and representative endpoint-specific configurations are then re-evaluated across multiple seeds against established baselines. This study therefore provides a reproducible benchmark structure for comparing future operator variants without confounding them with unrelated architectural changes, which can help identify which parts of the message-passing operator should receive priority when designing models for regression and classification tasks.

The remainder of the paper is structured as follows. Section 2 details the theoretical framework, experimental setup, and data analysis protocol; Section 3 reports operator-family effects, pairwise interactions, and baseline comparisons; Section 4 discusses implications and limitations; Section 5 concludes the study.

# 2. Methods

## 2.1 Theoretical framework

This section introduces the theoretical framework of the MPNN[8] and defines the specific message-construction and update operators evaluated in this benchmark.

### 2.1.1 Molecular Graph and Message Passing

We represent a molecule as an attributed graph $\mathcal{G} = (\mathcal{V}, \mathcal{E})$, where $\mathcal{V}$ denotes the set of atoms and $\mathcal{E}$ denotes the set of chemical bonds. Following common practice, each undirected chemical bond is treated as two directed edges[10]. Throughout this study, atom, edge, message-seed, and message representations are written as row vectors, and learnable linear maps are therefore right-multiplied. Each atom $i \in \mathcal{V}$ is associated with an $a$-dimensional row-vector representation $x_i^{(l)} \in \mathbb{R}^{1\times a}$ at layer $l$ (with initial features $x_i^{(0)}$), and each directed edge from atom $j$ to $i$ is associated with a $b$-dimensional row-vector feature $e_{ij} \in \mathbb{R}^{1\times b}$.

In an $L$-layer MPNN, node representations are iteratively updated through message passing. At each layer $l$, for a target atom $i$ and its neighboring source atoms $j \in \mathcal{N}(i)$, the message passing framework is defined as[8]:

$$m_{ij}^{(l)} = M^{(l)}\left(x_i^{(l)}, x_j^{(l)}, e_{ij}\right)$$

$$m_i^{(l)} = \sum_{j\in\mathcal{N}(i)} m_{ij}^{(l)}$$

$$x_i^{(l+1)} = U^{(l)}\left(x_i^{(l)}, m_i^{(l)}\right)$$

where $m_{ij}^{(l)}$ denotes the pairwise message generated by the message construction function $M^{(l)}(\cdot)$; $m_i^{(l)}$ represents the aggregated message, which is fixed in this benchmark as the sum of the pairwise messages, because summation is a permutation-invariant operation standard for aggregation in molecular MPNNs; and $U^{(l)}(\cdot)$ denotes the update operator.

### 2.1.2 Message Construction

We further decompose the message construction function $M^{(l)}(\cdot)$ into two sequential operations: message-seed initialization and node-edge fusion. At layer $l$, for each target atom $i$ and neighboring source atom $j \in \mathcal{N}(i)$, an intermediate pre-fusion representation termed message seed $s_{ij}^{(l)}$ is first generated from the node representations and subsequently integrated with the corresponding edge feature:

$$s_{ij}^{(l)} = I^{(l)}\left(x_i^{(l)}, x_j^{(l)}\right)$$

$$m_{ij}^{(l)} = F^{(l)}\left(s_{ij}^{(l)}, e_{ij}\right)$$

where $I^{(l)}(\cdot)$ denotes the message-seed initialization operator generating the message seed $s_{ij}^{(l)}$, and $F^{(l)}(\cdot)$ denotes the fusion operator merging the edge feature $e_{ij}$ with the message seed $s_{ij}^{(l)}$. For readability, the layer index superscript is often omitted, as in $s_{ij}$, $m_{ij}$, $m_i$, or $x_i$ when the layer is clear from context.

**2.1.2.1 Message-Seed Initialization**

Message-seed initialization herein refers exclusively to the generation of the message seed within each message-passing layer, distinct from the initial atom featurization $x_i^{(0)}$.

We evaluate four initialization operators (Init1–Init4):

**Init1 (Identity)** directly utilizes the neighboring source-node representation as the message seed:

$$s_{ij}^{(l)} = x_j^{(l)}.$$

**Init2 (Linear Projection)** applies a learnable linear transformation to the source-node representation:

$$s_{ij}^{(l)} = x_j^{(l)} W_a^{(l)},$$

where $W_a^{(l)}$ is a learnable weight matrix.

**Init3 (Degree Normalization)** introduces a structural inductive bias via GCN-style degree-normalized propagation[21]:

$$s_{ij}^{(l)} = \frac{x_j^{(l)}}{\sqrt{d_i d_j}},$$

where $d_i$ and $d_j$ represent the degrees of atoms $i$ and $j$, respectively.

**Init4 (Nonlinear Pairwise Transformation)** captures complex pairwise interactions between the target and source atoms through a nonlinear transformation:

$$s_{ij}^{(l)} = \mathrm{MLP}^{(l)}\left(x_i^{(l)} \parallel x_j^{(l)}\right),$$

where $\parallel$ denotes vector concatenation and $\mathrm{MLP}^{(l)}(\cdot)$ is a multilayer perceptron.

#### 2.1.2.2 Node-Edge Fusion

The message seed $s_{ij}^{(l)}$ is subsequently integrated with the edge feature $e_{ij}$ via node-edge fusion. Here we consider six edge-aware fusion operators alongside a no-edge reference condition:

**No-edge (None)** is the null reference, which omits explicit edge features, yielding $m_{ij}^{(l)} = s_{ij}^{(l)}$. This establishes a baseline to assess the contribution of edge attributes.

**Additive Fusion (Add)** projects the edge feature into the latent space of the message seed prior to element-wise addition:

$$m_{ij}^{(l)} = s_{ij}^{(l)} + e_{ij}W_e^{(l)},$$

where $W_e^{(l)}$ serves as the learnable projection matrix.

**Hadamard Fusion (Hadamard)** modulates the message seed via element-wise multiplication with the transformed edge representation:

$$m_{ij}^{(l)} = s_{ij}^{(l)} \odot \tanh(e_{ij}W_e^{(l)}),$$

where $\odot$ denotes the Hadamard product.

**Concatenation-Based Fusion (Concat1–4)** systematically explores the integration of edge features through concatenation. These four variants are distinguished by the presence of edge projection prior to concatenation and the application of a joint post-concatenation nonlinear transformation:

$$\text{Concat1:} \quad m_{ij}^{(l)} = s_{ij}^{(l)} \parallel e_{ij}$$

$$\text{Concat2:} \quad m_{ij}^{(l)} = \text{MLP}^{(l)}\left(s_{ij}^{(l)} \parallel e_{ij}\right)$$

$$\text{Concat3:} \quad m_{ij}^{(l)} = s_{ij}^{(l)} \parallel e_{ij}W_e^{(l)}$$

$$\text{Concat4:} \quad m_{ij}^{(l)} = \text{MLP}^{(l)}\left(s_{ij}^{(l)} \parallel e_{ij}W_e^{(l)}\right)$$

### 2.1.3 Node Update

We consider three update operators characterized by increasing transformation complexity.

**U1 (Linear Residual Update)** employs a residual linear combination of the current node state and the transformed aggregated message:

$$x_i^{(l+1)} = x_i^{(l)} + m_i^{(l)}W_m^{(l)},$$

where $W_m^{(l)}$ is a learnable weight matrix.

**U2 (Dual-Linear Update)** applies independent linear transformations to both the current node state and the aggregated message before summation:

$$x_i^{(l+1)} = x_i^{(l)} W_i^{(l)} + m_i^{(l)} W_m^{(l)},$$

where $W_i^{(l)}$ and $W_m^{(l)}$ are learnable weight matrices.

**U3 (GIN-style Nonlinear Update)** adopts the highly expressive update architecture of the Graph Isomorphism Network (GIN)[14]. The current node state and the aggregated message are combined and subsequently processed by an MLP:

$$x_i^{(l+1)} = \mathrm{MLP}^{(l)}\left((1+\epsilon) x_i^{(l)} + m_i^{(l)}\right),$$

where $\epsilon$ is fixed at zero following standard GIN implementations, a setting previously demonstrated to yield benchmark performance comparable to that of learnable $\epsilon$ configurations[14].

## 2.2 Experimental Setup

This section describes the experimental protocol for evaluating the operator design space, including the benchmark datasets, evaluation metrics, scaffold split, fixed-seed paired screening of the full operator space, representative multi-seed baseline comparisons, hyperparameter optimization procedure, and statistical analyses.

### 2.2.1 Operator Design Space

Table 1. Summary of the operator design space used to construct the 84 MPNN configurations.

| Operator family | Operators evaluated | Role in message passing | Dimension |
|---|---|---|---|
| Message-seed initialization | Init1, Init2, Init3, Init4 | Generates the pre-fusion message seed from source and/or target node states | 4 |
| Node-edge fusion | None, Add, Hadamard, Concat1, Concat2, Concat3, Concat4 | Incorporates bond features into the pairwise message | 7 |
| Node update | U1, U2, U3 | Updates node states after sum aggregation | 3 |

By orthogonally composing the operators defined in Sections 2.1.2 and 2.1.3, we construct a comprehensive benchmark design space comprising 84 unique MPNN configurations. This total includes 72 edge-aware combinations ($4 \times 6 \times 3$) and 12 additional no-edge reference conditions. Table 1 provides a summary of the operator families, and the full operator-design table is provided in the Supporting Information. We reiterate that the aggregation operation is fixed to summation for all configurations.

### 2.2.2 Benchmark Datasets

To comprehensively evaluate the proposed operators, we benchmarked our models on ten property prediction datasets from MoleculeNet[9], covering both classification and regression tasks (Table 2). For classification tasks, predictive performance is quantified using the area under the receiver operating characteristic curve (ROC-AUC)[22]. For regression, we report the root mean squared error (RMSE) for ESOL[23], FreeSolv[24], and Lipophilicity[9], and the mean absolute error (MAE) for the quantum chemical datasets, QM7[25] and QM8[26].

To rigorously assess out-of-distribution generalization, the datasets are partitioned using a scaffold-split procedure originally proposed by Wu *et al.*[9]. Specifically, molecules are clustered based on their two-dimensional Murcko frameworks generated via RDKit[27,28]. The data are subsequently divided into training, validation, and test sets according to an 8:1:1 ratio. For each dataset, the same scaffold split is reused across all benchmarked configurations and baseline models. This structure-aware splitting strategy provides a more challenging and realistic evaluation of model generalization to novel chemotypes compared to standard random splits.

Table 2. MoleculeNet benchmark datasets and primary evaluation metrics.

| Dataset | Task type | Property category | Metric* |
|---|---|---|---|
| ESOL | Regression | Aqueous solubility | RMSE (↓) |
| FreeSolv | Regression | Hydration free energy | RMSE (↓) |
| Lipophilicity | Regression | Octanol/water distribution coefficient | RMSE (↓) |
| QM7 | Regression | Quantum-chemical property | MAE (↓) |
| QM8 | Regression | Quantum-chemical property | MAE (↓) |
| BACE | Classification | Bioactivity | ROC-AUC (↑) |
| BBBP | Classification | Blood–brain barrier permeability | ROC-AUC (↑) |
| HIV | Classification | Antiviral activity | ROC-AUC (↑) |
| Tox21 | Classification | Toxicity | ROC-AUC (↑) |
| ClinTox | Classification | Clinical toxicity | ROC-AUC (↑) |

*Lower RMSE and MAE values (↓) indicate better regression performance, whereas higher ROC-AUC values (↑) indicate better classification performance.

### 2.2.3 Model Implementation, Training and Evaluation

Each of the 84 configurations, combined with fixed sum aggregation and sum readout, defines one MPNN model. Aggregation is fixed so that message-construction and update effects can be better isolated, whereas readout is fixed since it is a downstream graph-level pooling outside the message-passing design space. All 84 operator-space models are implemented with PyTorch Geometric (PyG)[29]. These models are compared to six representative GNN baselines, namely GIN[14], GCN[21], GAT[30], DMPNN[10], AttentiveFP[11], and Graphormer[3], in Section 3.3. GIN, GCN, GAT, AttentiveFP, and Graphormer are implemented with DGL[31](including DGL-LifeSci), whereas DMPNN uses the official Chemprop implementation. The baseline comparison preserves each baseline model's native readout mechanism, because that comparison evaluates complete molecular GNN architectures rather than isolating readout-controlled operator effects.

Table 3. Summary of the molecular graph featurization used in this benchmark.

| Graph component | Total dimensionality | Feature (dimensionality) |
|---|---|---|
| Atom | 133 | Atom type (101), number of bonds (7), formal charge (6), chirality (5), number of bonded hydrogens (6), hybridization (6), aromaticity (1), and atomic mass (1); |
| Bond | 14 | Null-bond indicator (1), bond type (4), conjugation (1), ring membership (1), and stereochemistry (7); |

Data processing, training, evaluation, and testing follow the Chemprop workflow[10]. SMILES strings are used as model input and converted into molecular graphs with RDKit[27] for feature initialization. This procedure generates 133-dimensional atom feature vectors and 14-dimensional bond feature vectors (Table 3). Among them, atomic mass is included as a scaled continuous feature, whereas the remaining features are categorical features and are represented using one-hot encoding. The full feature breakdown is provided in the Supporting Information.

To ensure reproducibility, the full 84-configuration benchmark uses a fixed scaffold split and parameter initialization seed of 0. For the representative baseline comparisons reported in Section 3.3, the selected operator configurations and baseline models are trained with seeds 0, 1 and 2.

Table 4. Hyperparameter search space used for Bayesian optimization across all operator configurations and baseline models.

| Hyperparameter | Search space | Sampling strategy |
|---|---|---|
| depth | {2, 3, 4, 5, 6} | Categorical |
| hidden_size | [128, 1024] | Quantized log-uniform (step = 32) |
| ffn_num_layers | {1, 2, 3} | Categorical |
| ffn_hidden_size | [128, 2048] | Quantized log-uniform (step = 32) |
| dropout | [0.0, 0.6] | Continuous uniform |

To ensure a fair comparison across the 84 configurations, hyperparameter tuning is standardized for all configurations. We use Bayesian optimization[32], implemented with the Hyperopt Python package[33] to search for optimal hyperparameters, including depth, hidden size, feed-forward network (FFN) layers and size, and dropout (detailed in Table 4). For each configuration–dataset pair, Bayesian optimization is run for 30 trials on the fixed split of that dataset and the selected hyperparameters are then used for the final model fit, and predictive performance is evaluated on the held-out test set using a fixed training seed of 0.

Models are trained for 100 epochs without early stopping to ensure convergence across the design space. Optimization follows the training protocol of Chemprop[10] and uses the Adam optimizer[34] with zero weight decay together with a Noam learning-rate scheduler[35]. The learning rate is increased during an initial warmup phase and then gradually decayed over the remaining steps, with an initial learning rate of $10^{-4}$.

Batch sizes are assigned according to dataset size: 32 for smaller benchmarks (BACE, BBBP, ESOL, FreeSolv, ClinTox), 64 for Lipophilicity, 128 for mid-sized datasets (QM7, Tox21), and 256 for the largest datasets (QM8, HIV).

Classification tasks use binary cross-entropy (BCE) loss, whereas regression tasks use mean squared error (MSE) loss. Training is performed on NVIDIA RTX 3090 GPUs, except for the HIV dataset, which is processed on NVIDIA V100 GPUs.

## 2.3 Data Analysis Protocol

To enable cross-dataset comparison across benchmarks with different metrics and scales, we standardize performance within each dataset for all 84 configurations. Raw scores as defined in Section 2.2.2 are converted into z-scores relative to the mean and standard deviation (SD) of the 84 configurations evaluated on that dataset. For regression tasks, more favorable configurations have lower such standardized z-scores; for classification tasks, more favorable configurations have higher z-scores. For each operator-family analysis, configurations sharing the same operator level are averaged within each dataset, producing a dataset-by-operator table for the relevant family. Operator-level summaries are then reported as mean $\pm$ SD across the five datasets in each endpoint family. The SD reported in the Std. score rows therefore reflects variation across datasets after within-dataset standardization, rather than variation across random seeds or repeated training runs.

Operator-family effects are evaluated using Friedman tests[36], with dataset treated as a repeated-measure block and regression and classification tasks analyzed separately. Statistical significance is assessed at $\alpha = 0.05$. Kendall's $W$[37] is reported as the Friedman effect size, reflecting the consistency of operator rankings across datasets. When the Friedman test indicates a significant family-level effect, post hoc pairwise comparisons are performed using exact two-sided Wilcoxon signed-rank tests[38] with Holm correction for multiple comparisons[39]. Pairwise operator interactions (e.g., initialization $\times$ fusion) are examined using exploratory two-way blocked ANOVA on standardized-score cell means, with dataset included as a blocking factor. The ANOVA summaries report $F$-statistics, corrected $p$-values, and partial $\eta^2$ values for main and interaction terms[40]. Because each endpoint family contains only five benchmark datasets, these interaction analyses are treated as exploratory complements to the nonparametric main-effect analyses rather than as standalone inferential evidence.

# 3. Results

This section first examines the effects of individual message construction and update operators, then analyzes their pairwise interactions, and finally compares competitive configurations in the design space with established molecular GNN baselines.

## 3.1 Effects of Individual Operator Components

Complete performance for all 84 configurations and their statistical analyses are provided in the Supporting Information. The complete heatmaps show that no single configuration dominates all datasets: the fixed-seed per-dataset optima differ across the ten benchmarks. This motivates the family-level analysis below, which asks which operator choices are repeatedly favorable across endpoint families rather than treating any one configuration as a universal winner. We first focus on the effects of message-seed initialization, node-edge fusion, and update operators as individual components of MPNN models.

### 3.1.1 Message-Seed Initialization

Tables 5 and 6 show the performance of the four message-seed initialization operators across regression and classification tasks, respectively. Initialization preferences differ between the two endpoint families. For regression, mean standardized scores are more favorable for the more complex Init3 and Init4 operators, with Init4 reaching the lowest mean standardized score (-0.388). For classification, mean standardized scores are more favorable for the simpler Init1 and Init2 operators, with Init2 reaching the highest mean standardized score (0.358).

Table 5. Regression performance averaged over configurations sharing the same message-seed initialization operator. Values are RMSE for ESOL, FreeSolv, and Lipophilicity and MAE for QM7 and QM8; lower values indicate better performance. Std. score denotes the mean ± standard deviation across datasets after within-dataset standardization, not across random seeds. Bold indicates the most favorable descriptive mean standardized score, not a statistically significant pairwise difference.

| Dataset | Init1 | Init2 | Init3 | Init4 |
|---|---|---|---|---|
| ESOL | 1.175 | 1.121 | 1.060 | 1.062 |
| FreeSolv | 2.201 | 2.179 | 2.050 | 1.912 |

| Dataset | Init1 | Init2 | Init3 | Init4 |
|---|---|---|---|---|
| Lipophilicity | 0.736 | 0.735 | 0.697 | 0.736 |
| QM7 | 91.300 | 88.800 | 88.900 | 77.000 |
| QM8 | 0.0183 | 0.0174 | 0.0165 | 0.0181 |
| Std. score | 0.418 ± 0.210 | 0.163 ± 0.172 | -0.193 ± 0.327 | **-0.388 ± 0.638** |

Friedman tests indicate significant family-level differences among initialization operators for both regression ($\chi_F^2(3) = 9.96, p = 0.0189$, Kendall's $W = 0.664$) and classification ($\chi_F^2(3) = 8.76$, $p = 0.0327$, $W = 0.584$). The Kendall's $W$ values indicate that initialization rankings are moderately to strongly consistent across the five datasets within each endpoint family. However, Holm-corrected Wilcoxon pairwise comparisons do not reach significance. Thus, the results support a family-level initialization effect in both regression and classification, with different descriptive directions, but they do not establish a definitive pairwise ranking among Init1–Init4.

Table 6. Classification performance averaged over configurations sharing the same message-seed initialization operator. Values are ROC-AUC; higher values indicate better performance. Std. score denotes the mean ± standard deviation across datasets after within-dataset standardization, not across random seeds. Bold indicates the most favorable descriptive mean standardized score, not a statistically significant pairwise difference.

| Dataset | Init1 | Init2 | Init3 | Init4 |
|---|---|---|---|---|
| BACE | 0.835 | 0.830 | 0.774 | 0.731 |
| BBBP | 0.915 | 0.921 | 0.901 | 0.898 |
| ClinTox | 0.781 | 0.783 | 0.762 | 0.777 |
| HIV | 0.742 | 0.746 | 0.749 | 0.736 |
| Tox21 | 0.836 | 0.836 | 0.828 | 0.831 |
| Std. score | 0.277 ± 0.254 | **0.358 ± 0.211** | -0.253 ± 0.288 | -0.382 ± 0.370 |

### 3.1.2 Node-Edge Fusion

Tables 7 and 8 show the performance of the seven node-edge fusion operators across regression and classification tasks, respectively. Fusion shows an even clearer regression-classification split than initialization. For regression, mean standardized scores are more favorable for MLP-based concatenation variants, with Concat4 reaching the lowest mean standardized score (-0.430) and Concat2 also performing favorably (-0.354). Add is the least favorable regression fusion operator (0.702). Classification shows a flatter and more dataset-dependent ranking: Hadamard has the highest mean standardized score (0.185), followed by Concat3 (0.130), whereas Add remains the least favorable on average (-0.391).

Friedman tests support a significant family-level fusion effect for regression ($\chi_F^2(6) = 23.06, p = 0.0008$, Kendall's $W = 0.769$), indicating that fusion rankings are relatively consistent across the five regression datasets. However, Holm-corrected

Wilcoxon pairwise comparisons do not identify any individually significant operator pair. Thus, within the significant family-level effect, the descriptive ranking is led by Concat4 and Concat2, but the analysis does not establish a definitive pairwise ordering among fusion operators. For classification, the Friedman test does not support an overall fusion effect ( $\chi_F^2(6) = 8.31$ , $p = 0.2160$ , $W = 0.277$ ), so the slight advantages of Hadamard and Concat3 should be interpreted as descriptive rather than statistically established.

Table 7. Regression performance averaged over configurations sharing the same node-edge fusion operator. Values are RMSE for ESOL, FreeSolv, and Lipophilicity and MAE for QM7 and QM8; lower values indicate better performance. Std. score denotes the mean ± standard deviation across datasets after within-dataset standardization, not across random seeds. Bold indicates the most favorable descriptive mean standardized score, not a statistically significant pairwise difference.

| Dataset | None | Add | Hadamard | Concat1 | Concat2 | Concat3 | Concat4 |
|---|---|---|---|---|---|---|---|
| ESOL | 1.109 | 1.237 | 1.085 | 1.116 | 1.035 | 1.123 | 1.028 |
| FreeSolv | 2.155 | 2.159 | 2.160 | 1.994 | 2.063 | 2.079 | 1.989 |
| Lipophilicity | 0.709 | 0.818 | 0.710 | 0.725 | 0.693 | 0.732 | 0.696 |
| QM7 | 88.400 | 89.800 | 87.700 | 87.100 | 82.900 | 87.100 | 82.600 |
| QM8 | 0.0174 | 0.0190 | 0.0176 | 0.0173 | 0.0170 | 0.0176 | 0.0169 |
| Std. score | 0.054 ± 0.207 | 0.702 ± 0.358 | 0.023 ± 0.199 | -0.059 ± 0.176 | -0.354 ± 0.191 | 0.063 ± 0.061 | **-0.430 ± 0.135** |

Table 8. Classification performance averaged over configurations sharing the same node-edge fusion operator. Values are ROC-AUC; higher values indicate better performance. Std. score denotes the mean ± standard deviation across datasets after within-dataset standardization, not across random seeds. Bold indicates the most favorable descriptive mean standardized score, not a statistically significant pairwise difference.

| Dataset | None | Add | Hadamard | Concat1 | Concat2 | Concat3 | Concat4 |
|---|---|---|---|---|---|---|---|
| BACE | 0.811 | 0.786 | 0.806 | 0.795 | 0.779 | 0.800 | 0.770 |
| BBBP | 0.892 | 0.905 | 0.907 | 0.917 | 0.921 | 0.910 | 0.911 |
| ClinTox | 0.797 | 0.741 | 0.786 | 0.768 | 0.776 | 0.774 | 0.787 |
| HIV | 0.738 | 0.738 | 0.751 | 0.733 | 0.746 | 0.749 | 0.749 |
| Tox21 | 0.833 | 0.829 | 0.834 | 0.833 | 0.832 | 0.836 | 0.831 |
| Std. score | -0.007 ± 0.539 | -0.391 ± 0.391 | **0.185 ± 0.181** | -0.037 ± 0.315 | 0.085 ± 0.290 | 0.130 ± 0.143 | 0.035 ± 0.285 |

The contrast between the favorable descriptive performance of Concat4 in regression and its weaker classification trend motivates a targeted ablation study. We examined whether the full Concat4 design, particularly its edge projection and post-concatenation MLPs, is necessary across endpoint families. We therefore performed a component ablation with fixed message-seed initialization and node update, using Init1 and U1.

The analysis is performed on four representative MoleculeNet datasets: ESOL and Lipophilicity for regression, and BACE and BBBP for classification. This choice covers both endpoint families examined in this study while keeping the ablation computationally tractable. We constructed three simplified variants by removing the edge projection, the post-concatenation MLP, or both components from the full Concat4. Specifically, C4-minimal removes both components; C4-noProj removes only the edge projection; and C4-noMLP removes only the post-concatenation MLP.

The ablation results on representative regression and classification datasets are shown in Table 9. The experiments follow the same fixed scaffold split and training protocol as the full 84 configurations, and the reported values are test-set results obtained with seed 0. On the two regression datasets (ESOL and Lipophilicity), full Concat4 yields the lowest error, and removing either the edge projection or the post-concatenation MLP increases error, most visibly on ESOL. On the two classification datasets (BACE and BBBP), C4-minimal attains the highest ROC-AUC. Because this ablation uses one initialization-update setting and only four datasets, it should be viewed as supportive case evidence rather than a standalone conclusion. Even so, it is consistent with the broader pattern that the regression tasks examined here benefit more from multi-stage edge transformation than the classification tasks do.

Table 9. Targeted component ablation of the Concat4 fusion operator on representative regression and classification datasets. Values are RMSE for ESOL and Lipophilicity and ROC-AUC for BACE and BBBP.

| Variant | ESOL | Lipophilicity | BACE | BBBP |
|---|---|---|---|---|
| C4-minimal | 1.642 | 0.981 | 0.875 | 0.939 |
| C4-noProj | 1.017 | 0.977 | 0.861 | 0.936 |
| C4-noMLP | 1.399 | 0.975 | 0.863 | 0.933 |
| C4-full | 0.967 | 0.650 | 0.830 | 0.938 |

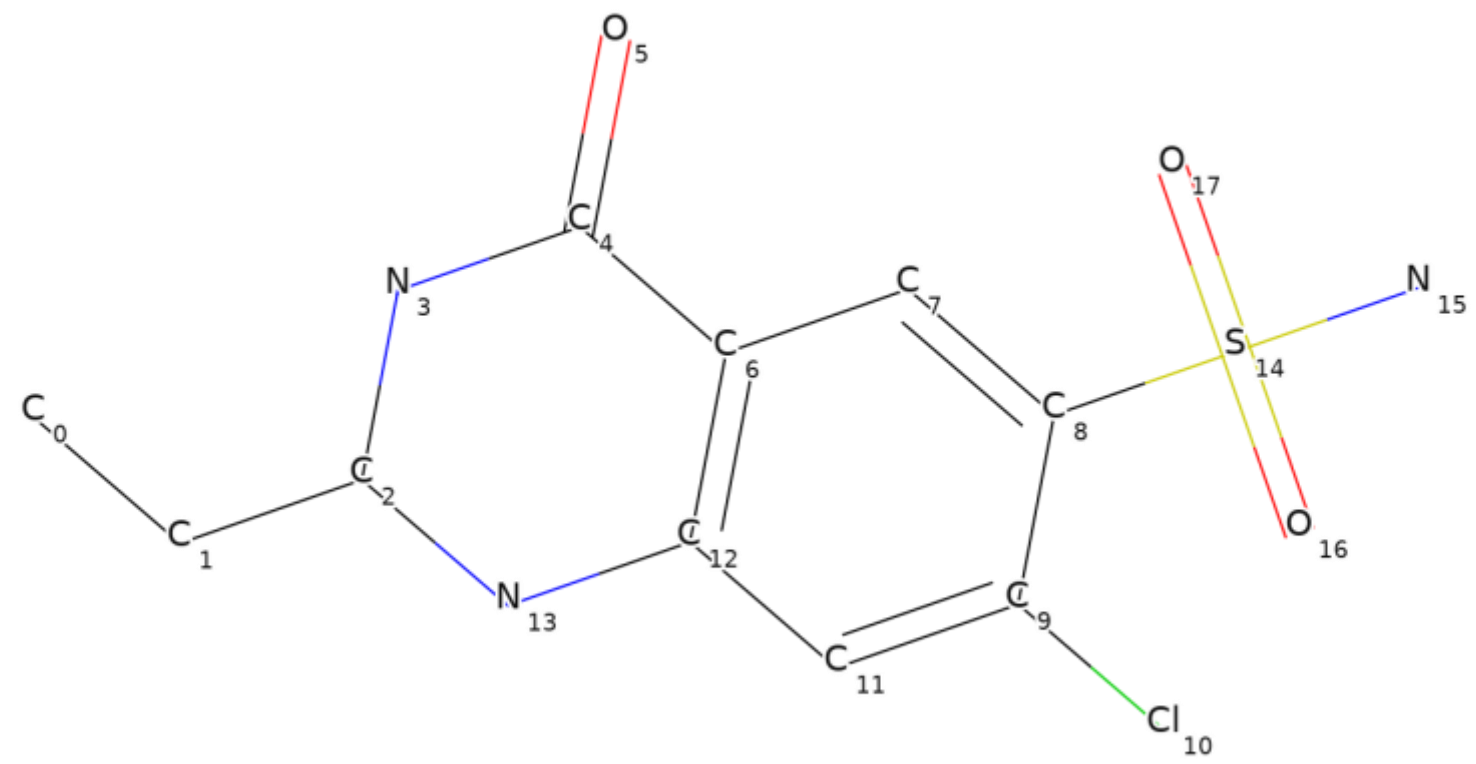

Figure 1. Atom-indexed molecular graph for the Quinethazone molecule. Labels follow RDKit 0-based atom indexing and correspond to the notation used in the subsequent distance comparisons.

While the benchmark and ablation results suggest different fusion preferences for regression and classification, they do not reveal how fusion operators affect the internal representations of individual molecules. To visually probe this atom-level behavior, we compared Concat4 and Hadamard on the Quinethazone molecule (CAS: 73-49-4; SMILES: CCC2NC(=O)c1cc(c(Cl)cc1N2)S(N)(=O)=O)[41], whose atom-indexed molecular graph is shown in Figure 1. This molecule contains multiple functional groups, including a sulfonamide, a carbonyl, and an aromatic ring, providing a useful context for examining how different fusion operators shape atom-level representations across message-passing layers. We focus on the latent representation of the ring nitrogens ($N_3$ and $N_{13}$), the sulfonamide nitrogen ($N_{15}$), the carbonyl oxygen ($O_5$), and the sulfonyl oxygens ($O_{16}$ and $O_{17}$). These atoms include symmetry-related pairs and chemically distinct heteroatoms with identical or similar initial features, making them informative probes for how fusion operators modulate representation geometry in functionally relevant contexts.

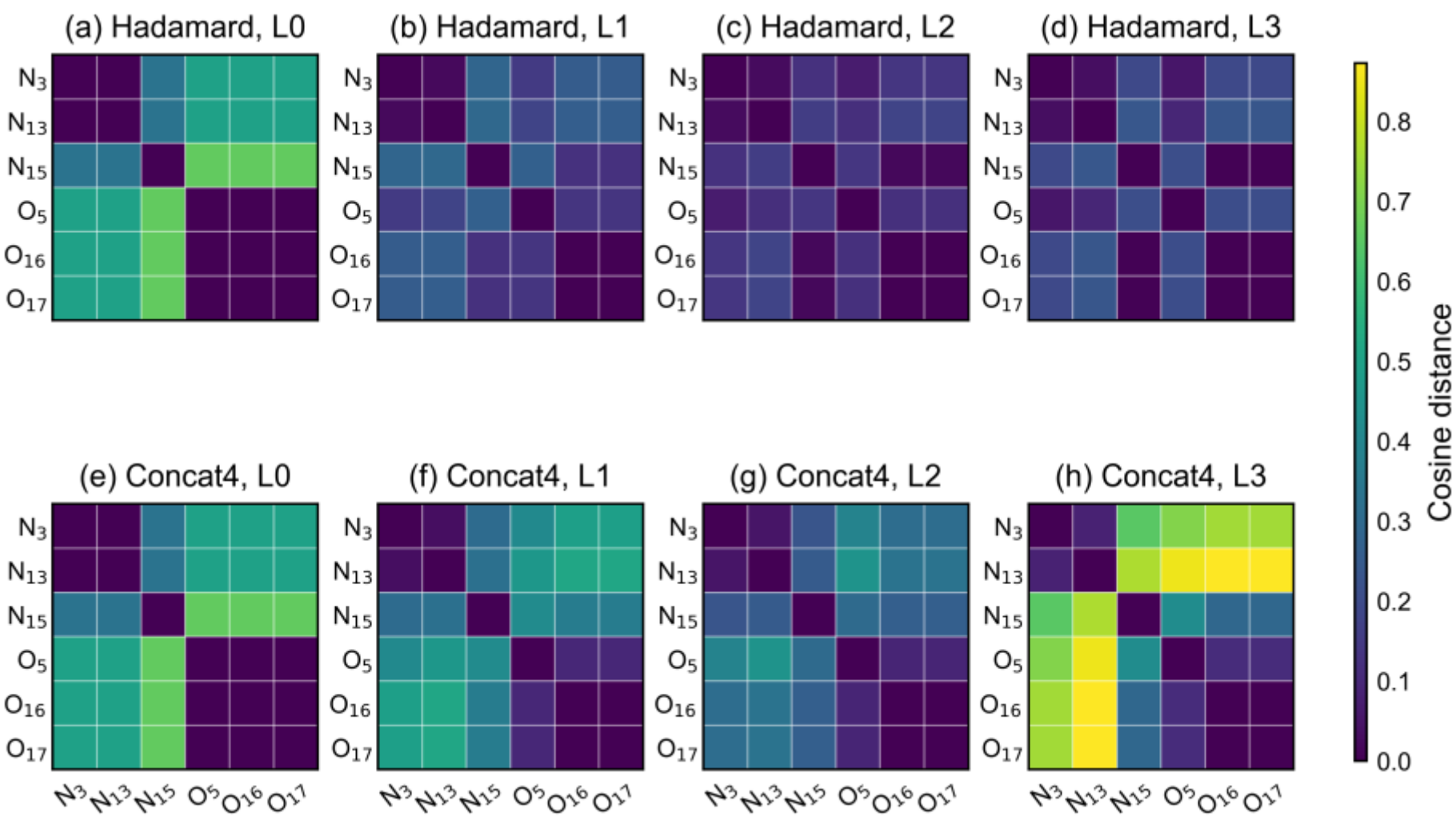

Figure 2. Pairwise cosine-distance heatmaps of oxygen (brown) and nitrogen (blue) atom representations in the full learned representation space of the Quinethazone molecule.

Figure 2 shows pairwise cosine-distance heatmaps of the representations of selected N and O atoms in Quinethazone across message-passing layers. At layer 0, Hadamard and Concat4 share the same initial feature-derived distance structure: the sulfonyl oxygens ($O_{16}$ and $O_{17}$) have zero distance, and the ring nitrogens ($N_3$ and $N_{13}$) have near-zero distance. These symmetry-related relationships are preserved during message passing in both models. The oxygen atoms also form a low-distance block that is separated from the ring nitrogens and the sulfonamide nitrogen.

The two fusion operators differ in how they modulate the distances among these atom groups across layers. Hadamard reduces the distances between the nitrogen and oxygen groups in layer 1 and maintains a relatively compact geometry throughout. In contrast, although Concat4 also slightly reduces these distances in layer 1, it increases the distances between the groups at layer 2, creating clearer separation between the nitrogen and oxygen atoms. This pattern suggests that Concat4 may facilitate stronger feature-space remixing and differentiation among functionally distinct atom groups than Hadamard.

In both models, the layer-3 heatmaps show reduced separation among selected heteroatom groups compared with the layer-2 heatmaps, indicating a degree of oversmoothing at the final layer. However, Concat4 still retains substantially larger final-layer distances than Hadamard. In the final Concat4 heatmap, the ring-nitrogen pair ($N_3$ and $N_{13}$) remains strongly separated from $N_{15}$ and from the oxygen atoms ($O_5$, $O_{16}$, and $O_{17}$), while the near-zero distances within the $N_3$/$N_{13}$ and $O_{16}$/$O_{17}$ pairs are preserved.

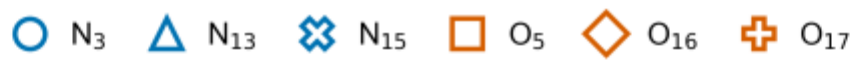

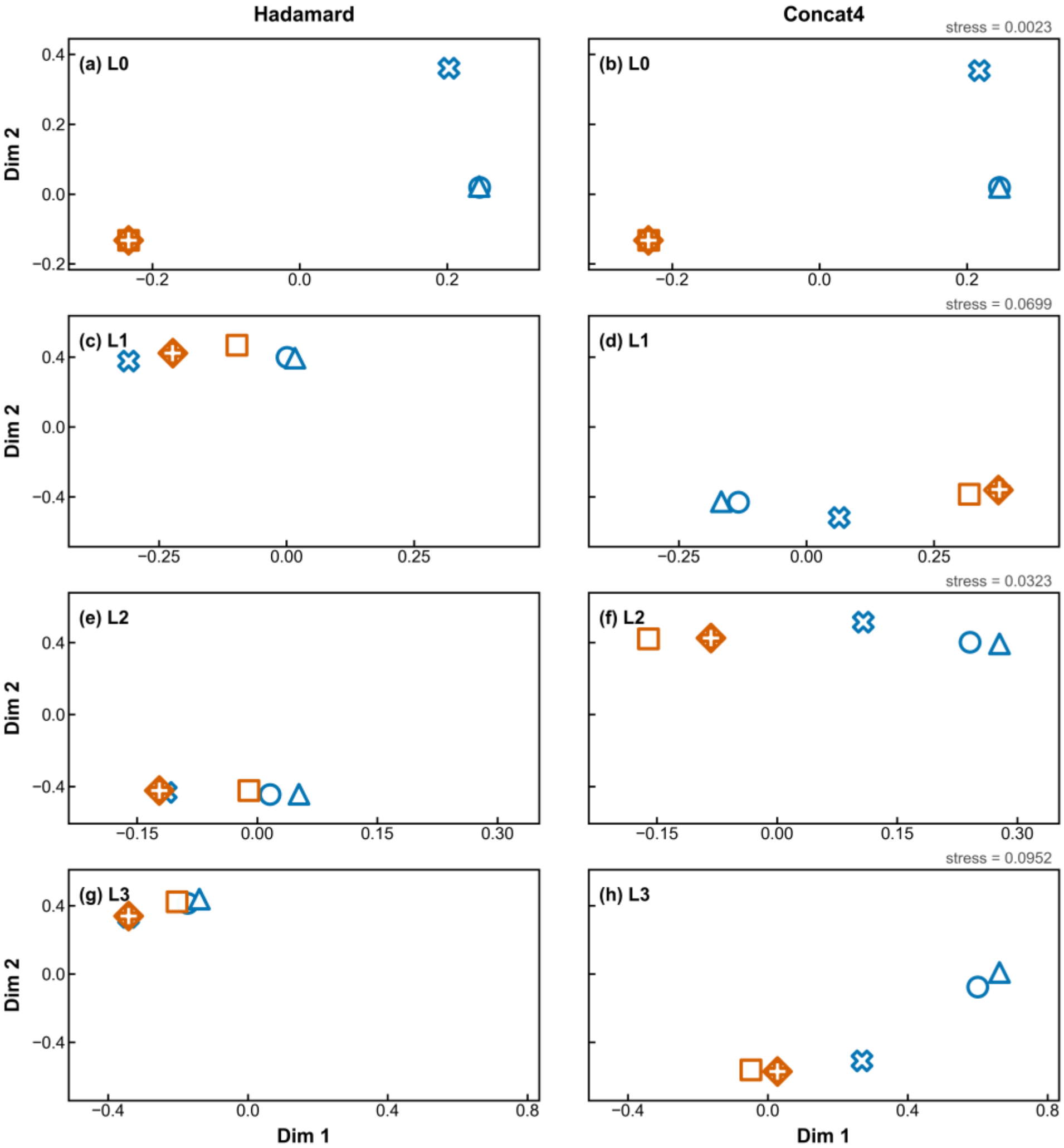

Figure 3. Layer-wise joint multi-dimensional scaling (MDS) projections of oxygen (brown) and nitrogen (blue) atom representations obtained in the Hadamard and Concat4 fusion-based models for the Quinethazone molecule. The atom indices follow those in Figure 1.

The joint multi-dimensional scaling (MDS) projections[42] (Figure 3) are consistent with this pattern: Concat4 preserves a more separated final-layer geometry, whereas Hadamard collapses the selected atom representations into a more compressed arrangement. Thus, this case study suggests that Concat4 is less prone to final-layer oversmoothing than Hadamard, consistent with its stronger feature-space remixing capacity.

### 3.1.3 Node Update

Tables 10 and 11 show the performance of the three node-state update operators across regression and classification tasks. The effect of update is weaker and less stable than the effects of message-seed initialization and node-edge fusion. In regression, U3 obtains the lowest mean standardized score (-0.288), followed by U2 (0.039), suggesting a descriptive advantage for the MLP-based update. In classification, all three update operators are close to zero in mean standardized score, with U1, U2, and U3 differing mainly by small dataset-specific fluctuations.

Table 10. Regression performance averaged over configurations sharing the same node-state update operator. Values are RMSE for ESOL, FreeSolv, and Lipophilicity and MAE for QM7 and QM8; lower values indicate better performance. Std. score denotes the mean ± standard deviation across datasets after within-dataset standardization, not across random seeds. Bold indicates the most favorable descriptive mean standardized score, not a statistically significant pairwise difference.

| Dataset | U1 | U2 | U3 |
|---|---|---|---|
| ESOL | 1.146 | 1.114 | 1.054 |
| FreeSolv | 2.056 | 2.057 | 2.144 |
| Lipophilicity | 0.765 | 0.727 | 0.687 |
| QM7 | 88.100 | 87.500 | 83.900 |
| QM8 | 0.0183 | 0.0177 | 0.0166 |
| Std. score | 0.248 ± 0.216 | 0.039 ± 0.092 | -0.288 ± 0.282 |

Table 11. Classification performance averaged over configurations sharing the same node-state update operator. Values are ROC-AUC; higher values indicate better performance. Std. score denotes the mean ± standard deviation across datasets after within-dataset standardization, not across random seeds.

| Dataset | U1 | U2 | U3 |
|---|---|---|---|
| BACE | 0.794 | 0.775 | 0.809 |
| BBBP | 0.909 | 0.911 | 0.906 |
| ClinTox | 0.765 | 0.777 | 0.786 |
| HIV | 0.744 | 0.752 | 0.734 |
| Tox21 | 0.834 | 0.832 | 0.832 |
| Std. score | -0.017 ± 0.186 | 0.023 ± 0.242 | -0.006 ± 0.294 |

Friedman tests do not support a statistically significant update-family effect for either regression ($\chi_F^2(2) = 3.60, p = 0.1653$, Kendall's $W = 0.360$) or classification ($\chi_F^2(2) = 0.00, p = 1.000, W = 0.000$). Thus, no formal pairwise ranking is inferred among U1, U2, and U3. The apparent regression advantage of U3 remains descriptive rather than statistically established, and update choice appears less influential than message-seed initialization or node-edge fusion under the present protocol.

## 3.2 Pairwise Operator Interactions

The individual-operator analyses above identify message construction as the main source of performance variation, but they do not show whether the operator families act independently. We therefore use the pairwise interaction analyses as a bridge between the marginal results in Section 3.1 and the representative full-model comparisons in Section 3.3. The goal is not to rank every cell in the factorial design. Rather, it is to determine whether the favorable choices in one operator family remain favorable across the levels of another, and where apparent gains depend on a specific pairing.

### 3.2.1 Initialization–Fusion Interactions

Initialization and fusion interact most clearly in the regression benchmarks. In the blocked ANOVA on within-dataset standardized scores, both marginal terms are significant (initialization: $F(3{,}388) = 14.48$, $p = 5.8 \times 10^{-9}$, $\eta_p^2 = 0.101$; fusion: $F(6{,}388) = 8.65$, $p = 7.7 \times 10^{-9}$, $\eta_p^2 = 0.118$), and the initialization × fusion interaction is also supported ($F(18{,}388) = 3.23$, $p = 1.18 \times 10^{-5}$, $\eta_p^2 = 0.130$). Thus, the regression advantage of a fusion operator is conditional on the message-seed initialization with which it is paired.

The regression heatmap makes this conditionality concrete (Figure 4). Concat4 is the most favorable fusion choice under Init1 and Init2, Concat2 is most favorable under Init3, and Concat1 becomes most favorable under Init4. The global best cell is Init4 + Concat1, whereas Init1 + Add is the least favorable among the averaged standardized scores. This pattern refines the marginal result in Section 3.1.2: concatenation-based fusion is generally favorable for regression, but no single concatenation variant dominates across all initialization regimes. Regression performance is therefore better described as a coordinated message-construction effect than as two fully independent main effects.

The classification-side pattern is weaker and should be interpreted more cautiously. The initialization main effect remains strong ($F(3{,}388) = 14.53$, $p = 5.5 \times 10^{-9}$, $\eta_p^2 = 0.101$), whereas the fusion main effect ($F(6{,}388) = 2.14$, $p = 0.0477$, $\eta_p^2 = 0.032$) and initialization × fusion interaction ($F(18{,}388) = 1.83$, $p = 0.0207$, $\eta_p^2 = 0.078$) are smaller. The heatmap is consistent with this hierarchy (Figure 5): favorable cells are concentrated around Init1 and Init2, with Init2 + Concat4 and Init1 + Concat3 giving the highest averaged standardized scores, while several Init3 and Init4 pairings are unfavorable. Because the classification fusion family was not significant in the Friedman analysis (Section 3.1.2) and only five classification datasets are available, these cell-level differences should be read as exploratory. The robust conclusion is that classification is more sensitive to message-seed initialization than to a stable fusion preference.

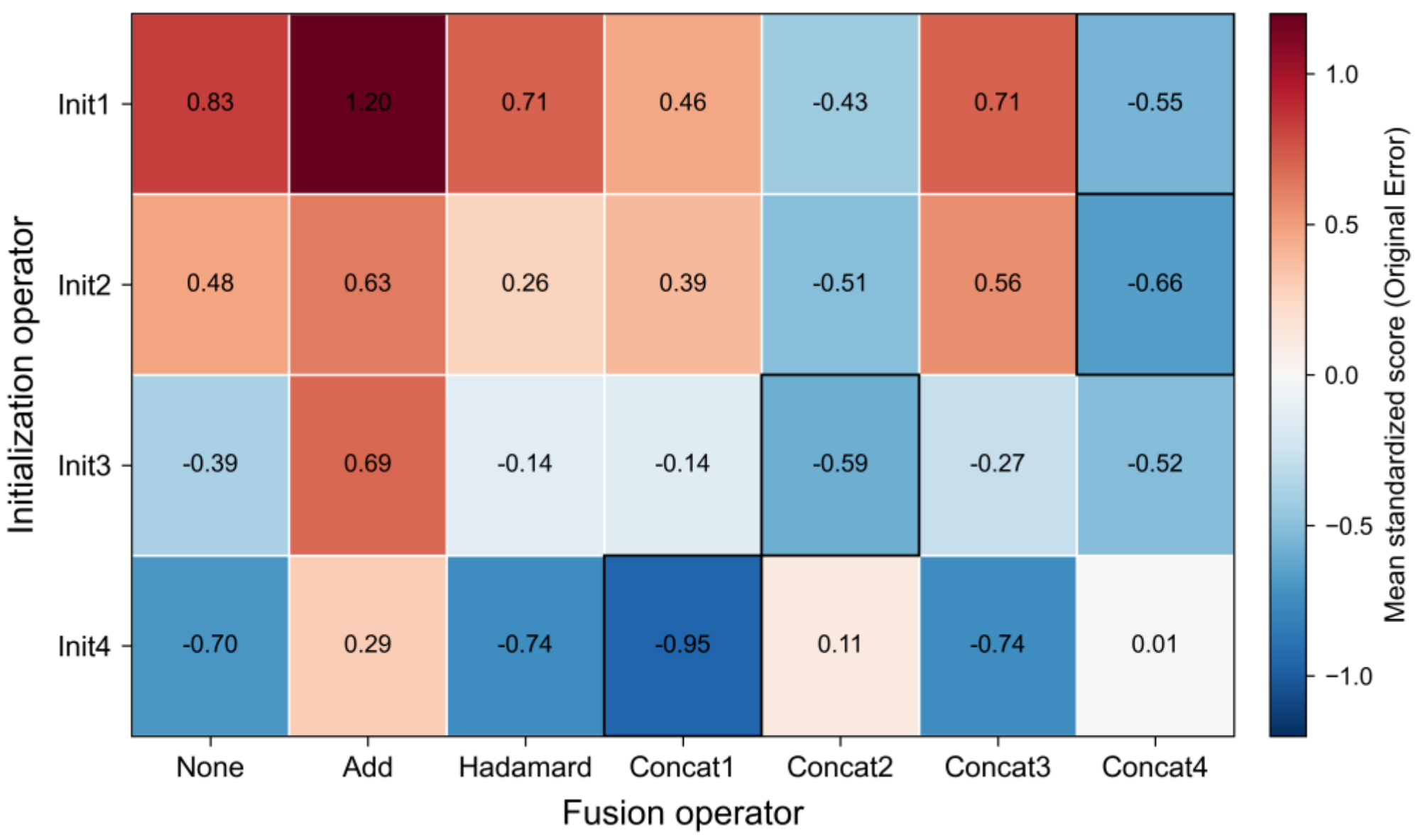

Figure 4. Initialization–fusion interaction patterns for regression tasks. Each cell shows the mean within-dataset standardized score for an initialization–fusion combination, averaged across the five regression benchmarks. Lower scores indicate better performance; black borders mark the most favorable fusion operator within each initialization group.

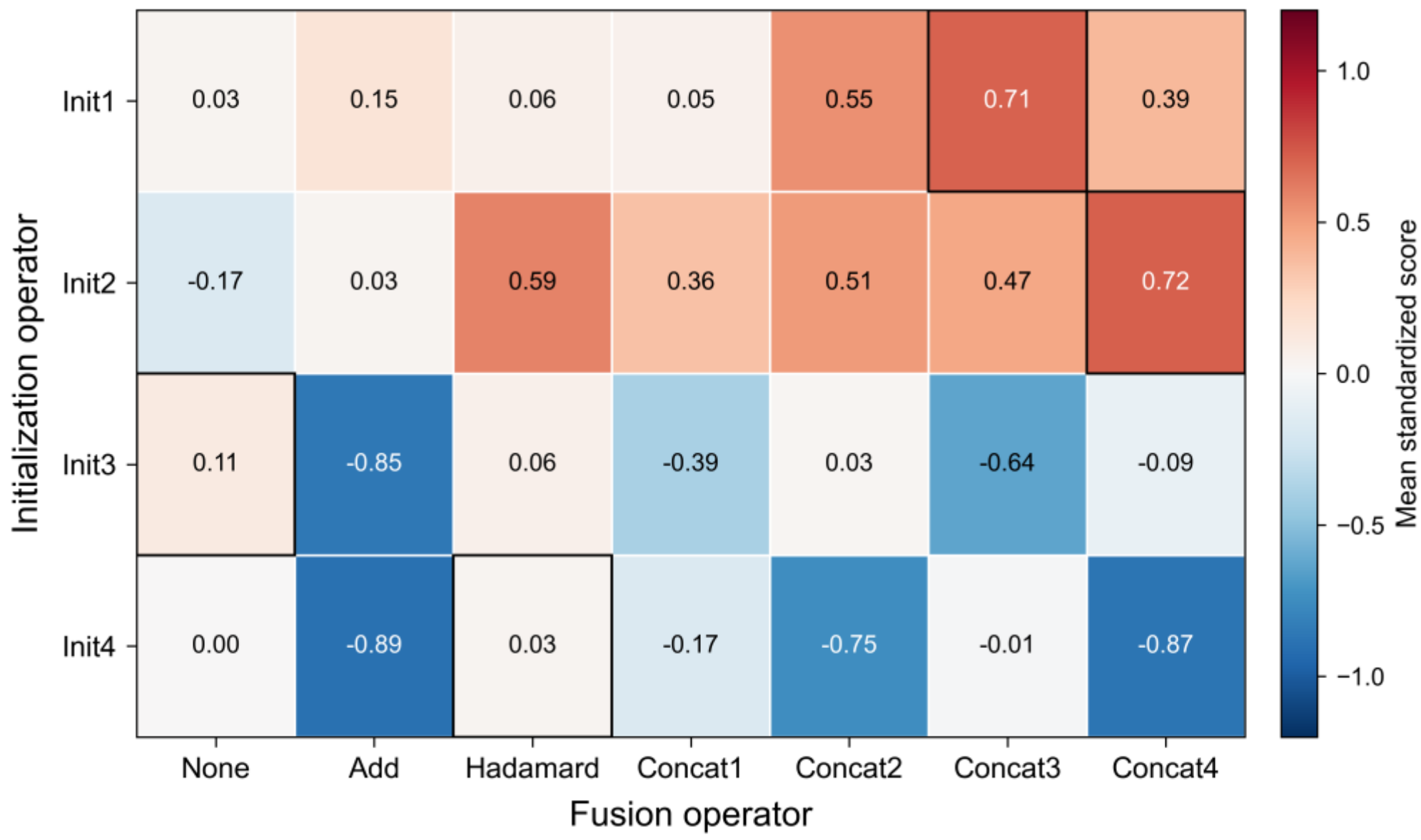

Figure 5. Initialization–fusion interaction patterns for classification tasks. Each cell shows the mean within-dataset standardized score for an initialization–fusion combination, averaged across the five classification benchmarks. Higher scores indicate better performance; black borders mark the most favorable fusion operator within each initialization group.

### 3.2.2 Fusion–Update Interactions

The second interaction analysis asks whether the update operator changes the fusion ranking after messages have been constructed and aggregated. Here the evidence for coupling is weak. In regression, the fusion × update interaction is not statistically supported ($F(12{,}395) = 0.39$, $p = 0.966$, $\eta_p^2 = 0.012$), even though the blocked ANOVA gives significant marginal terms for fusion ( $F(6{,}395) = 7.75$ , $p =$ $6.84 \times 10^{-8}$, $\eta_p^2 = 0.105$) and update ($F(2{,}395) = 9.73$, $p = 7.49 \times 10^{-5}$, $\eta_p^2 = 0.047$). The latter update term should be treated cautiously because the nonparametric Friedman analysis in Section 3.1.3 did not support an update-family effect for regression ($p = 0.1653$).

The regression heatmap therefore reads less like an interaction pattern and more like a marginal fusion pattern with a modest update-side shift (Figure 6). Concat4 is most favorable under U1 and U2, Concat2 is most favorable under U3, and Add remains unfavorable across all three updates. U3 often lowers the standardized score relative to U1 and U2, consistent with its descriptive regression advantage in Section 3.1.3, but this shift does not reorganize the fusion landscape. In practical terms, choosing a nonlinear update cannot compensate for an unfavorable fusion operator in the way that a compatible initialization-fusion pairing can improve message construction.

Classification shows even less evidence for fusion-update structure. Neither the fusion main effect ($F(6{,}395) = 2.04$, $p = 0.0594$, $\eta_p^2 = 0.030$), the update main effect ($F(2{,}395) = 0.06$, $p = 0.940$, $\eta_p^2 < 0.001$), nor the fusion × update interaction ($F(12{,}395) = 0.46$, $p = 0.939$, $\eta_p^2 = 0.014$) reaches conventional significance. The classification heatmap contains descriptively favorable cells such as Hadamard + U1 and Concat3 + U1 (Figure 7), but these local differences do not form a supported interaction. Taken together, the pairwise analyses localize meaningful operator coupling to the message-construction stage, especially for regression, and support treating update choice as a secondary tuning dimension under the present benchmark.

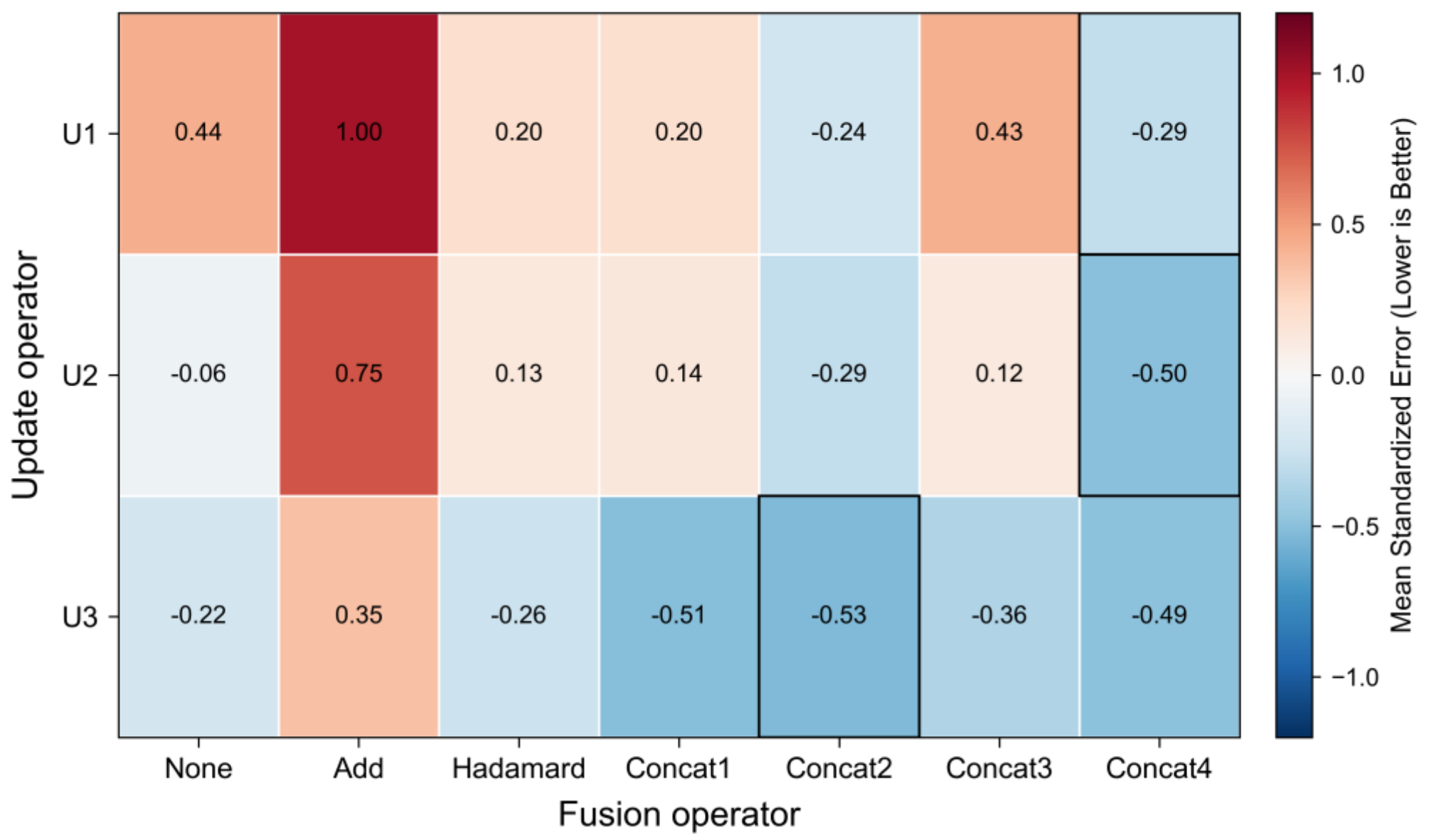

Figure 6. Fusion–update interaction patterns for regression tasks. Each cell shows the mean within-dataset standardized score for a fusion–update combination, averaged across the five regression

benchmarks. Lower scores indicate better performance; black borders mark the most favorable fusion operator within each update group.

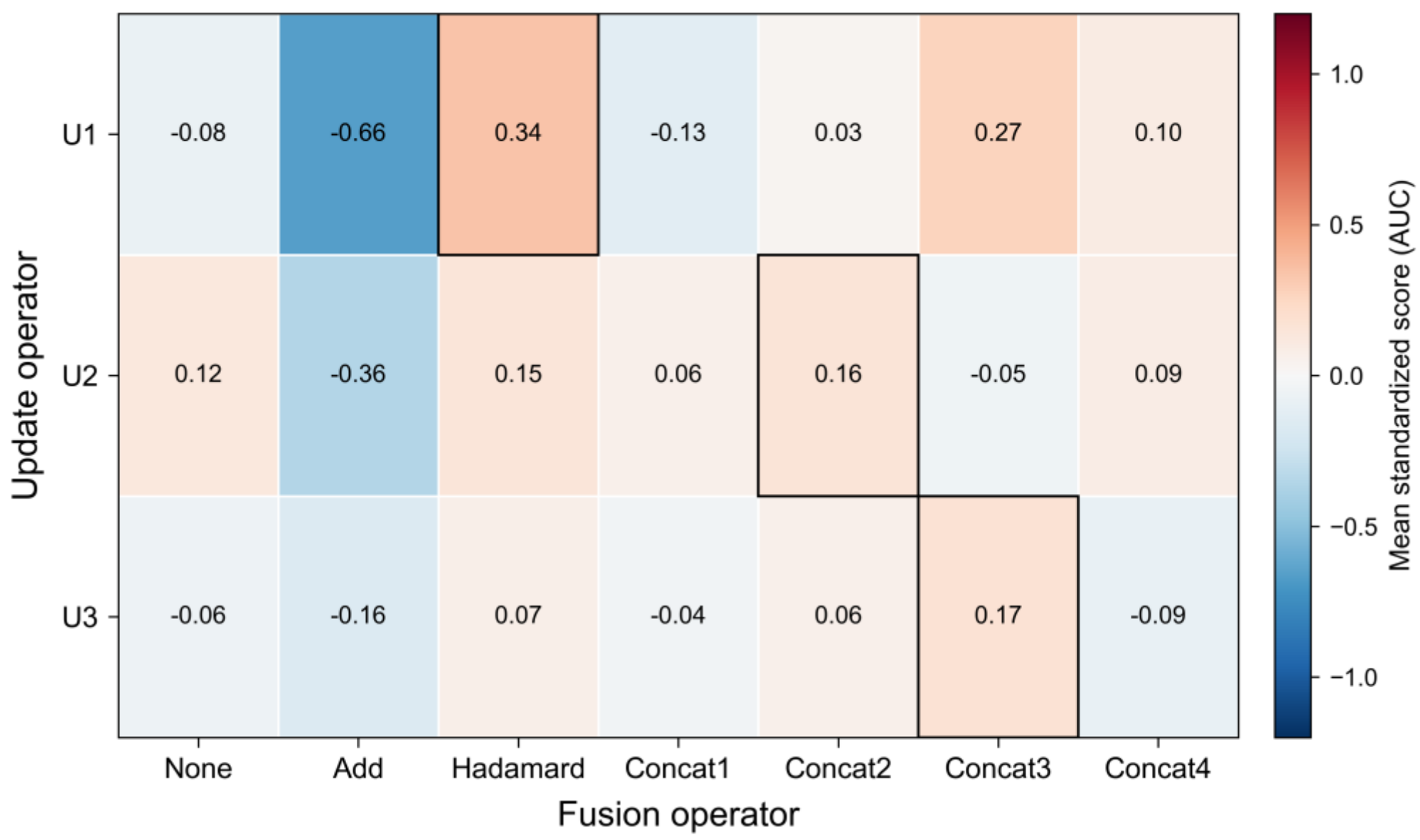

Figure 7. Fusion–update interaction patterns for classification tasks. Each cell shows the mean within-dataset standardized score for a fusion–update combination, averaged across the five classification benchmarks. Higher scores indicate better performance; black borders mark the most favorable fusion operator within each update group.

## 3.3 Comparison with Molecular GNN Baselines

After examining the effects of individual operators and their interactions, we next ask whether configurations selected separately for regression and classification can recover competitive performance relative to established molecular GNN baselines. We therefore compare the strongest representative configuration from each endpoint family under the same scaffold-split, tuning, and evaluation protocol. For classification, Init2 + Concat4 + U1 attains the highest mean raw ROC-AUC across the five classification benchmarks; for regression, where the disparate scales of RMSE and MAE motivate within-dataset standardization, Init4 + Concat1 + U2 attains the lowest mean standardized score across the five regression benchmarks. This comparison provides a controlled reference for the performance reachable within the decomposed operator space under a shared tuning budget.

Table 12. Held-out test performance of the representative classification configuration Init2 + Concat4 + U1 and molecular GNN baselines. Values are ROC-AUC and are reported as mean ± standard deviation across three runs with random seeds 0, 1 and 2; higher values indicate better performance.

| Model | HIV | Tox21 | ClinTox | BACE | BBBP |
|---|---|---|---|---|---|
| GIN | 0.728 ± 0.015 | 0.792 ± 0.015 | 0.782 ± 0.015 | 0.818 ± 0.020 | 0.881 ± 0.074 |
| GCN | 0.727 ± 0.007 | 0.816 ± 0.015 | 0.792 ± 0.015 | 0.803 ± 0.013 | 0.883 ± 0.005 |
| GAT | 0.718 ± 0.004 | 0.808 ± 0.016 | 0.801 ± 0.006 | 0.812 ± 0.023 | 0.857 ± 0.019 |

| Model | HIV | Tox21 | ClinTox | BACE | BBBP |
|---|---|---|---|---|---|
| DMPNN | 0.757 ± 0.007 | 0.836 ± 0.002 | 0.831 ± 0.068 | 0.838 ± 0.004 | 0.921 ± 0.039 |
| AttentiveFP | 0.751 ± 0.021 | 0.837 ± 0.008 | 0.825 ± 0.015 | 0.838 ± 0.021 | 0.902 ± 0.040 |
| Graphormer | 0.737 ± 0.012 | 0.822 ± 0.001 | 0.693 ± 0.076 | 0.835 ± 0.029 | 0.917 ± 0.042 |
| This work | 0.765 ± 0.008 | 0.841 ± 0.003 | 0.839 ± 0.008 | 0.847 ± 0.011 | 0.928 ± 0.010 |

Table 13. Held-out test performance of the representative regression configuration Init4 + Concat1 + U2 and molecular GNN baselines. Values are RMSE for Lipophilicity, FreeSolv, and ESOL and MAE for QM7 and QM8, reported as mean ± standard deviation across three runs with random seeds 0, 1, and 2; lower values indicate better performance.

| Model | Lipophilicity | FreeSolv | ESOL | QM7 | QM8 |
|---|---|---|---|---|---|
| GIN | 0.742 ± 0.082 | 2.628 ± 0.604 | 1.314 ± 0.331 | 93.776 ± 4.282 | 0.0176 ± 0.0004 |
| GCN | 0.754 ± 0.100 | 2.472 ± 0.737 | 1.197 ± 0.198 | 88.380 ± 1.248 | 0.0181 ± 0.0001 |
| GAT | 0.731 ± 0.056 | 2.399 ± 0.463 | 1.213 ± 0.021 | 87.228 ± 1.001 | 0.0177 ± 0.0002 |
| DMPNN | 0.644 ± 0.001 | 2.074 ± 0.453 | 1.051 ± 0.093 | 85.644 ± 4.101 | 0.0155 ± 0.0001 |
| AttentiveFP | 0.644 ± 0.051 | 2.106 ± 0.533 | 1.036 ± 0.088 | 68.006 ± 6.728 | 0.0146 ± 0.0020 |
| Graphormer | 1.050 ± 0.026 | 3.566 ± 0.190 | 1.044 ± 0.165 | 69.265 ± 9.847 | 0.0144 ± 0.0020 |
| This work | 0.632 ± 0.005 | 1.768 ± 0.288 | 0.908 ± 0.152 | 76.021 ± 1.016 | 0.0154 ± 0.0002 |

Table 12 shows that the Init2 + Concat4 + U1 configuration achieves the highest numerical ROC-AUC among the compared models on all five classification benchmarks. Although the performance margin is modest on some datasets, the representative operator combination remains numerically competitive throughout the evaluation. In contrast, Table 13 presents a more heterogeneous pattern for regression tasks. The Init4 + Concat1 + U2 configuration achieves the best performance on Lipophilicity, FreeSolv, and ESOL, whereas the lowest errors for QM7 and QM8 are attained by AttentiveFP and Graphormer, respectively. Taken together, these results suggest that operator combinations selected separately for regression and classification can reach performance comparable to, and in several cases numerically better than, strong molecular GNN baselines under a unified protocol. The pattern also differs by endpoint family: classification shows a more consistent numerical trend, whereas regression exhibits greater variation across datasets. Because these representative configurations were selected post hoc, these comparisons should be interpreted as descriptive rather than inferential. Nevertheless, the emergence of different numerically favorable combinations for classification and regression is consistent with the operator-family patterns examined in the preceding sections.

# 4. Discussion

The benchmark separates three operator families that are usually bundled inside named MPNN architectures. Four observations organize the discussion: initialization effects are detectable in both endpoint families, fusion effects are clearest for regression, update-family effects are comparatively weak under the present protocol, and post hoc representative configurations remain competitive with established molecular GNN baselines. These observations are interpreted as empirical design heuristics rather than universal architectural rules.

## 4.1 Message Construction as the Primary Driver

Under fixed sum aggregation, operations upstream of aggregation—message-seed initialization and node-edge fusion—are associated with more consistent performance differences than the subsequent node-update step. The Friedman tests support initialization-family effects for both endpoint families and a fusion-family effect for regression, whereas the update-family tests do not reach significance. Within a sum-aggregation molecular MPNN design, this pattern suggests a practical search order: evaluate initialization and fusion choices before allocating substantial effort to update-operator complexity.

This conclusion should be read empirically. Expressive update functions remain theoretically important in GNN analysis[14], and the present benchmark does not test all molecular settings. Rather, within this fixed-aggregation, 2D molecular design space, the measured performance differences are more consistently associated with the construction of pairwise messages than with the transformation applied after aggregation.

## 4.2 Regression-Classification Differences in Fusion

Node-edge fusion is where the regression-classification split becomes most visible. In the regression benchmarks, the fusion family has a statistically supported effect, and the descriptive ranking favors Concat4 and Concat2. In the classification benchmarks, the fusion ranking is flatter: Hadamard and Concat3 are numerically favorable, but the family-level Friedman test does not support a reliable overall effect. This difference suggests that the useful role of bond features depends not only on whether edges are used, but also on how strongly edge features are allowed to reshape atom-derived messages.

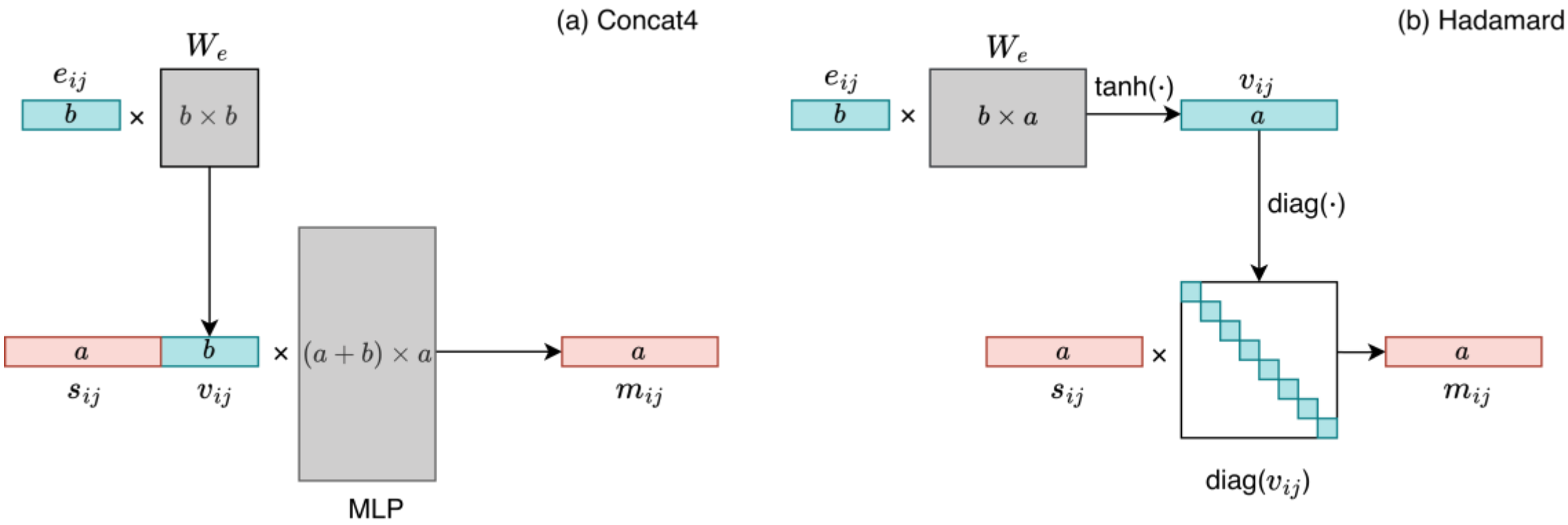

Figure 8. Node-edge fusion mechanisms. (a) Concat4 projects the edge feature $e_{ij}$ to an edge embedding $v_{ij}$, concatenates it with the atom-derived message seed shown as $x_j$, and passes the joint vector through an MLP to produce the pairwise message $m_{ij}$. This path can learn dense nonlinear interactions between atom and bond channels. (b) Hadamard fusion projects $e_{ij}$ to a same-dimensional gate $v_{ij} = \tanh(e_{ij}W_e)$, represents the gate as the diagonal operator $\mathrm{diag}(v_{ij})$, and rescales the message seed channel by channel to produce $m_{ij}$. The contrast is therefore between nonlinear node-edge remixing in Concat4 and edge-conditioned channel modulation in Hadamard.

The mechanism schematic illustrates one possible interpretation (Figure 8). Hadamard fusion acts as an edge-conditioned channel gate on the message seed, whereas Concat4 applies a nonlinear mixer to the message seed and a projected edge representation. This distinction provides a compact explanation for the regression trend: concatenation-based mixing can represent denser atom-bond interactions, while gating provides a more conservative modulation of existing latent channels.

The evidence for this interpretation is suggestive rather than definitive. It is supported by the regression fusion main effect, the initialization–fusion interaction pattern, the Concat4 ablation, and the Quinethazone representation probe; however, the classification fusion family is not significant, and the representation analysis is a single-molecule case study. Therefore, the practical conclusion is limited: richer fusion is an empirically useful regression-oriented heuristic in this benchmark, not a universal preference for all molecular endpoints. Additional mechanistic detail is provided in the Supporting Information.

## 4.3 Update Complexity: When Does Nonlinearity Help?

The update results sharpen the preceding point: once a message has been constructed and summed, adding expressive nonlinear transformations at the node-update stage is not automatically decisive. U3 has the most favorable descriptive mean standardized score in regression, but the update-family effect is not statistically supported for either endpoint family, and the fusion × update interaction is weak.

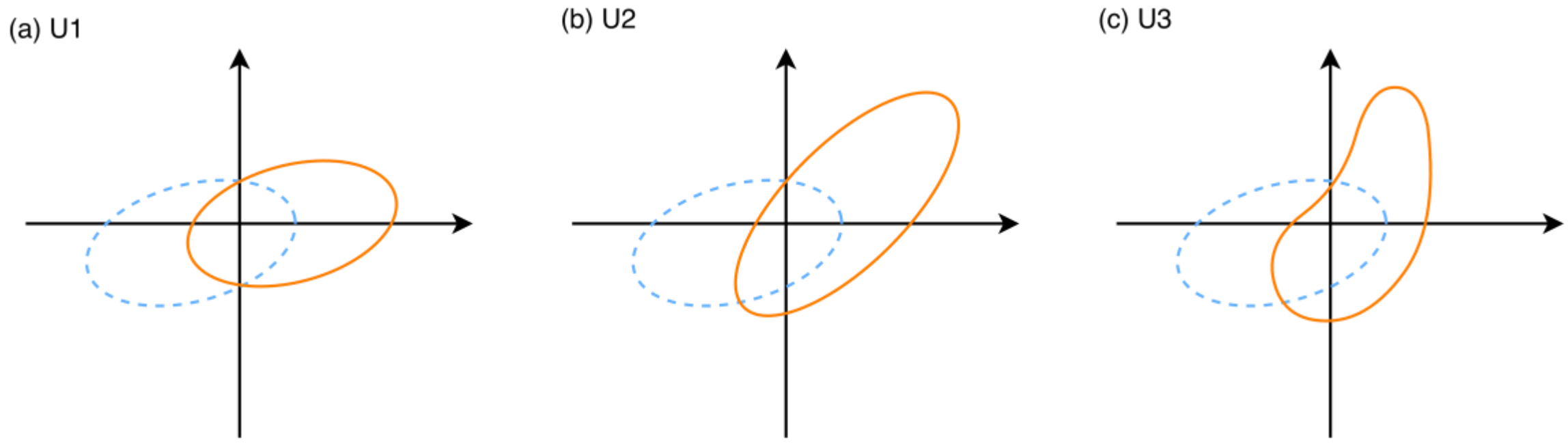

Figure 9. Geometric interpretation of the node-state update operators. The blue dashed contour indicates the current node-state region, and the orange contour indicates the region after incorporating the aggregated message. (a) U1 is a residual update: the current state keeps an identity path, while the message contributes an additive linear displacement, so the original coordinate frame is largely preserved. (b) U2 applies learned linear maps to both the current state and the message before addition, allowing rotation, scaling, and shearing of the representation space. (c) U3 applies an MLP to the combined state-message signal, replacing a fixed affine map with a nonlinear transformation whose local geometry can vary across the space. The panels illustrate increasing transformation flexibility rather than an empirical ranking of U1, U2, and U3.

Figure 9 summarizes a restrained geometric interpretation. U1 preserves an identity path and adds a transformed message, U2 applies learned linear maps to both terms, and U3 applies a nonlinear transformation after the self-state and aggregated message are combined. Because all three updates act after fixed sum aggregation, they can reshape the aggregated neighborhood summary but cannot recover pairwise atom-bond details that are not captured during message construction.

The central empirical claim is therefore limited. Nonlinear update expressiveness is not a universal requirement under the present benchmark, and any benefit of U3 appears secondary to message construction. The update-family Friedman tests do not reach significance in either endpoint family (regression $p = 0.1653$; classification $p = 1.000$), so the geometric account should be read as an interpretation of the observed pattern, not as proof of a general mechanism. Additional update-operator interpretation is provided in the Supporting Information.

## 4.4 Operator Coupling and Baseline Recovery

Two additional observations sit at the interface between operator choice and whole-model comparison and are worth considering together. First, the exploratory blocked-ANOVA analysis in Section 3.2.1 suggests that, for regression tasks, the effect of a fusion operator depends on the message-seed initialization operator with which it is paired, whereas classification tasks show a much weaker pattern in which initialization carries the main signal and the additional fusion and interaction terms should be treated as exploratory. By contrast, the fusion × update analysis in Section 3.2.2 does not support coupling in either endpoint family, consistent with the weaker update main effect; the message-construction stage is therefore the locus of both the strongest operator-family effects and the clearest pairwise coupling in this benchmark. This pattern is consistent with the broader observation that regression endpoints in this benchmark are more sensitive to how the message is constructed, and it cautions against treating initialization and fusion as fully independent tuning dimensions for regression.

Second, the reference comparison in Section 3.3 shows that, under the same implementation, evaluation protocol, and hyperparameter-optimization budget, representative regression and classification configurations drawn from this design space reach numerical performance comparable to established molecular GNN baselines on eight of the ten datasets. Because the representative configurations were selected post hoc and the comparison was not designed as a formal hypothesis test, these numbers should be read descriptively rather than as a ranking claim. Even so, the result is useful: carefully chosen 2D operator configurations can be competitive with more specialized architectures under a shared protocol. This reinforces the view that operator-level factorial benchmarking should accompany whole-model leaderboard comparisons in molecular MPNN studies.

## 4.5 Limitations and Future Directions

Our experimental design is intended to isolate operator-level behavior, and this controlled setting defines the scope of the conclusions. The analysis is limited to 2D molecular graphs, fixed sum aggregation, fixed sum readout, and a representative set of message-construction and update operators. This design reduces confounding from geometry, attention mechanisms, readout variants, and other architectural choices, but it also means that the observed preferences should not be assumed to transfer directly to 3D conformational tasks or to equivariant and invariant architectures in which spatial interactions are central. In addition, although MoleculeNet provides standardized benchmarks, it does not capture the full heterogeneity, assay bias, label noise, endpoint diversity, and training-test redundancy encountered in medicinal chemistry, toxicology, and materials discovery[9,43,44]. Moreover, the datasets included in the present benchmark are not large by modern deep-learning standards, with most tasks falling in the thousand- to ten-thousand-sample range. As a result, although the current study is informative for small- to medium-scale molecular benchmarks, the behavior of these models and operator choices on substantially larger datasets remains uncertain and should be evaluated explicitly in future work. A related statistical caveat is that, with only five datasets per endpoint family, the Friedman and Holm-corrected Wilcoxon tests have limited power to resolve pairwise orderings; the recurring pattern of significant family-level effects accompanied by non-significant Holm-corrected pairs in Section 3.1 should therefore be read as power-limited rather than as evidence that operators are interchangeable within a family.

Given the extensive combinatorial space evaluated (84 operator variants across 10 datasets), the full benchmarking matrix is executed under a standardized scaffold split and fixed random seed (seed = 0) to preserve paired comparability across configurations while keeping the benchmark computationally tractable. This protocol establishes a consistent basis for within-benchmark relative comparisons and nonparametric analyses of operator-level trends, but it does not quantify variability arising from alternative scaffold partitions, weight initializations, or independent training runs. The focused baseline comparisons in Section 3.3 partially address this issue by re-evaluating selected representative configurations and baseline models across three seeds (0, 1, and 2), but the full operator space has not yet been benchmarked under multi-seed evaluation. Consequently, future work should build on these controlled observations by extending multi-seed evaluation and repeated scaffold splits to a broader subset of high-performing configurations. Such extensions, alongside the exploration of conformation-aware 3D model families and broader datasets, would further clarify which structural preferences reflect generalizable principles for matching graph neural network operators to downstream chemical tasks. Finally, we believe the conclusions should largely hold even when aggregation and readout are not held fixed, but future research should validate this claim.

# 5. Conclusion

As molecular representation learning advances toward geometric and attention-based paradigms, rigorous understanding of the 2D message-passing GNNs remains essential. We present an operator-level factorial benchmark of molecular MPNNs, spanning 84 configurations across ten MoleculeNet tasks under a unified scaffold-split, matched-tuning, and statistical-analysis protocol. Within this scope, the clearest family-level effects arise before aggregation: message-seed initialization is supported for both endpoint families, node-edge fusion is supported for regression, and update-family choice is not statistically supported under the present protocol.

The practical implication is an empirical search heuristic rather than a universal design rule. Practitioners can first examine the 4×7 initialization-fusion space before tuning update complexity, reducing the first-pass architectural search from 84 to 28 configurations. In the post hoc baseline comparison, two configurations selected separately for regression and classification ranked numerically best on eight of ten benchmark datasets under the shared protocol, supporting the value of operator-level screening as a complement to whole-model leaderboard comparison.

The main contribution is therefore not a single MPNN architecture, but a reproducible way to evaluate which operator families matter for a given endpoint setting. Future work should extend this framework to repeated scaffold splits, broader multi-seed evaluation, additional operator families, and 3D molecular representations.

## Data Availability

The source codebase is available on GitHub: https://github.com/AI4Mater/MPNN-Operator-Benchmark. Experimental data (both raw and processed) are archived on Zenodo: https://doi.org/10.5281/zenodo.20279020.

## Supporting Information

Detailed operator-design definitions, Chemprop featurization details, baseline implementation details, complete performance heatmaps for all 84 operator configurations across the 10 MoleculeNet datasets, supplementary statistical summaries for Friedman tests and blocked two-way ANOVA analyses, supplementary mechanistic interpretation, and reproducibility notes.

## Author Information

### Corresponding Author

Wei Xie - Materials Genome Institute, Shanghai University, Shanghai 200444, China; orcid.org/0000-0003-1501896X; Email: xiewei@xielab.org

### Authors

Panyu Jiao - Materials Genome Institute, Shanghai University, Shanghai 200444, China

Shuizhou Chen - School of Computer Engineering and Science, Shanghai University, Shanghai 200444, China

Yiheng Shen - Materials Genome Institute, Shanghai University, Shanghai 200444, China

Yuyang Wang - Materials Genome Institute, Shanghai University, Shanghai 200444, China

Runhai Ouyang - School of Materials Science and Engineering, Tongji University, Shanghai 201804, China

## Author Contributions

[#]P.J, [#]S.C. and [#]Y.S. contribute equally to this work. P.J.: conception, implementation, experiment, data analysis, writing; S.C.: conception, data analysis, writing; Y.S.: conception, data analysis, supervision; Y.W.: implementation; O.R.: supervision; W.X.: conception, data analysis, writing, supervision, funding acquisition.

## Notes

The authors declare no competing financial interest.

## Acknowledgments

This work was supported by National Natural Science Foundation of China (52003150), the Program for Young Eastern Scholar at Shanghai Institutions of Higher Education (QD2019006), the Science and Technology Commission of Shanghai Municipality (No.25511103400 and 24CL2901702), and Shanghai Technical Service Computing Center of Science and Engineering, Shanghai University.